\documentclass{ecai} 
\usepackage{latexsym}
\usepackage{amssymb}
\usepackage{amsmath}
\usepackage{amsthm}
\usepackage{booktabs}
\usepackage{enumitem}
\usepackage{graphicx}
\usepackage{color}
\usepackage{amsmath,amsfonts}
\usepackage{algorithmic}
\usepackage{algorithm}
\usepackage{array}
\usepackage[caption=false,font=normalsize,labelfont=sf,textfont=sf]{subfig}
\usepackage{textcomp}
\usepackage{stfloats}
\usepackage{url}
\usepackage{verbatim}
\usepackage{graphicx}
\usepackage{tikz}
\usepackage{booktabs}
\usepackage{verbatim}
\usepackage[framemethod=default]{mdframed}
\usepackage[utf8]{inputenc}
\usepackage{pgfplots}

\usepackage{adjustbox}

\pgfplotsset{compat=1.16}

\usetikzlibrary{arrows.meta, positioning, shapes.geometric, fit, backgrounds, calc, shapes.multipart, arrows}
\newcommand{\BibTeX}{B\kern-.05em{\sc i\kern-.025em b}\kern-.08em\TeX}

\begin{document}

%%%%%%%%%%%%%%%%%%%%%%%%%%%%%%%%%%%%%%%%%%%%%%%%%%%%%%%%%%%%%%%%%%%%%%%%

\begin{frontmatter}

%%% Use this command to specify your submission number.
%%% In doubleblind mode, it will be printed on the first page.

\paperid{123} 

%%% Use this command to specify the title of your paper.

\title{Document Retrieval Augmented Fine-Tuning ({\it DRAFT}) for safety-critical software assessments}

%%% Use this combinations of commands to specify all authors of your 
%%% paper. Use \fnms{} and \snm{} to indicate everyone's first names 
%%% and surname. This will help the publisher with indexing the 
%%% proceedings. Please use a reasonable approximation in case your 
%%% name does not neatly split into "first names" and "surname".
%%% Specifying your ORCID digital identifier is optional. 
%%% Use the \thanks{} command to indicate one or more corresponding 
%%% authors and their email address(es). If so desired, you can specify
%%% author contributions using the \footnote{} command.

\author[A]{\fnms{Regan}~\snm{Bolton}\thanks{Corresponding Author}}
\author[B]{\fnms{Mohammadreza}~\snm{Sheikhfathollahi}}
\author[B]{\fnms{Simon}~\snm{Parkinson}} 
\author[A]{\fnms{Vanessa}~\snm{Vulovic}} 
\author[A]{\fnms{Gary}~\snm{Bamford}} 
\author[A]{\fnms{Dan}~\snm{Basher}} 
\author[A]{\fnms{Howard}~\snm{Parkinson}}

\address[A]{Digital Transit Limited, 3M Buckley Innovation Centre, UK,  HD1 3BD}
\address[B]{Department of Computer Science, University of Huddersfield, UK, HD1 3DH}

%%% Use this environment to include an abstract of your paper.

\begin{abstract}
Safety critical software assessment requires robust assessment against complex regulatory frameworks, a process traditionally limited by manual evaluation. This paper presents {\bf D}ocument {\bf R}etrieval-{\bf A}ugmented {\bf F}ine-{\bf T}uning (DRAFT), a novel approach that enhances the capabilities of a large language model (LLM) for safety-critical compliance assessment. DRAFT builds upon existing Retrieval-Augmented Generation (RAG) techniques by introducing a novel fine-tuning framework that accommodates our dual-retrieval architecture, which simultaneously accesses both software documentation and applicable reference standards. To fine-tune DRAFT, we develop a semi-automated dataset generation methodology that incorporates variable numbers of relevant documents with meaningful distractors, closely mirroring real-world assessment scenarios. Experiments with GPT-4o-mini demonstrate a 7\% improvement in correctness over the baseline model, with qualitative improvements in evidence handling, response structure, and domain-specific reasoning. DRAFT represents a practical approach to improving compliance assessment systems while maintaining the transparency and evidence-based reasoning essential in regulatory domains.
\end{abstract}

\end{frontmatter}

%%%%%%%%%%%%%%%%%%%%%%%%%%%%%%%%%%%%%%%%%%%%%%%%%%%%%%%%%%%%%%%%%%%%%%%%

\section{Introduction}

Systems running safety-critical software operate in domains where failures can have severe consequences, including loss of life, environmental damage, or significant financial losses~\cite{knight2002safety, leveson2016engineering}. Ensuring these systems are developed according to rigorous safety standards requires comprehensive assessment methodologies that can effectively evaluate complex documentation against established regulatory frameworks~\cite{kornecki2009certification}. Traditional assessment approaches are often manual, time-consuming and subject to human error, creating the need for automated solutions that can maintain the necessary level of accuracy and traceability~\cite{rushby2009software}.
Recent applications of large language models (LLMs) have demonstrated their potential for document analysis tasks~\cite{brown2020language}, with Retrieval-Augmented Generation (RAG) emerging as a promising approach to enhance LLM capabilities with external knowledge~\cite{lewis2020retrieval, gao2023retrieval}. However, while RAG excels at retrieving and integrating information, its effectiveness in domain-specific tasks such as the assessment of safety-critical software documentation remains limited by several factors: imperfect retrieval processes, difficulty distinguishing between relevant and irrelevant information, and challenges in maintaining clear evidence traceability that is essential in regulatory contexts~\cite{zhao2402retrieval, barnett2024seven}.

Fine-tuning LLMs for specialised domains offers an alternative approach~\cite{patil2024review}, but conventional fine-tuning methods often struggle with maintaining the model's ability to use retrieved information effectively and may lead to overfitting or catastrophic forgetting~\cite{zhang2024scaling, balaguer2024rag}. Combining RAG with fine-tuning presents additional challenges, as the interaction between these methods is not straightforward and can sometimes lead to reduced performance~\cite{soudani2024fine, lakatos2024investigating}.

In this paper, we propose an adaptation of Retrieval-Augmented Fine-Tuning (RAFT)~\cite{zhang2024raft} specifically designed for the assessment of safety-critical software documentation. Our approach leverages an existing dual-retrieval architecture~\cite{cyrail_paper} that simultaneously accesses both documentation and applicable standards, and focusses on fine-tuning a model to work effectively within this architecture. We specifically improve assessment by training the model to process and respond to a comprehensive set of compliance queries derived from industry standards, allowing systematic assessment against regulatory requirements.

Our approach, which we refer to as {\bf D}ocument {\bf R}etrieval-{\bf A}ugmented {\bf F}ine-{\bf T}uning ({\it DRAFT}), addresses the unique requirements of safety-critical software process assessment by:
\begin{enumerate}
\item Developing a semi-automated dataset generation methodology integrating our dual-retriever architecture.
\item Implement a fine-tuning framework that promotes selective information use while maintaining direct citation and traceability.
\item Training models to effectively differentiate between relevant and irrelevant information while optimising for domain-specific reasoning in compliance contexts.
\end{enumerate}

We demonstrate the effectiveness of our approach through experiments with GPT-4o-mini models~\cite{hurst2024gpt}, showing significant improvements in the accuracy and robustness of compliance assessment to irrelevant information. Our methodology provides a pragmatic solution to improve the assessment of safety-critical software while maintaining the transparency and evidence-based reasoning required in regulatory domains.

The remainder of this paper is organised as follows. In Section~\ref{sec:appcontext}, we present the application context and motivation. Section \ref{sec:related_work} reviews related work in RAG, fine-tuning and RAFT; Section \ref{sec:methodology} presents our methodology in detail, Section \ref{sec:results} discusses our results, Section \ref{sec:discussion} discusses trends and observations, and finally, Section \ref{sec:conclusion} concludes with a discussion of implications and future directions.

\section{Application context}
\label{sec:appcontext}
%{\bf The Role of an Assessor in Safety-Critical Software Development}

One of the key domains for safety-critical software is the railway industry. As the railway domain is increasingly digitalised, the need for software as part of systems from rolling stock to signalling has increased. As part of the supply chain, software is required to be developed to a Software Integrity Level (SIL) rating. SIL ratings range between 1 and 4 which are defined by probability of failure. SIL 1, the lowest, will generally have lower software safety requirements than SIL 4. These SIL levels are defined in the Euronorm standard EN50716 and represent increasing levels of rigour in the development process to prevent systematic errors.

The standard also defines the role of an assessor. ``The Assessor shall be independent of the project team and shall be a different entity, organisationally independent, from those undertaking other roles in the project''. The role of assessment is different from that of verification. Rather than individually reviewing every document for correctness, the assessor instead makes a higher-level judgement call as to whether the standard has been followed satisfactorily. The assessor must be satisfied that all appointed personnel can demonstrate competency in their roles, that the software is fit for its intended purpose, and that the activities outlined in the standard have been carried out to a sufficient level, meaning they have been performed with appropriate rigour, documentation, and verification relative to the required SIL. This process is both knowledge- and time-intensive, increasing the cost and constraining the assessment process. 

Furthermore, an assessor may carry out audits and inspections, for example, test witnessing, throughout the development process. The results of all these activities are recorded and summarised in a Software Assessment Report, alongside any nonconformities and a final judgement. Non-conformities might include inadequate traceability from requirements to design, insufficient test coverage, or lack of evidence for specific verification activities. This Software Assessment Report provides confidence from the customer and regulators that the process defined by the standard has been followed correctly, giving credence to the SIL level achieved.

In this research, the purpose of the application is to provide the assessor with an automated tool to evaluate safety-critical software. By allowing direct queries against documentation, it aids the assessment process, making report creation more efficient and thorough.

\section{Related work}
\label{sec:related_work}

{\bf Retrieval-Augmented Generation (RAG)}
\newline
% RAG has emerged as one of the most advanced AI techniques for enhancing large language models (LLMs) by integrating external knowledge sources~\cite{zhao2402retrieval}, providing reliable and up-to-date external knowledge, and offering significant convenience for a variety of tasks~\cite{fan2024survey,wang2024searching}. 
RAG has become one of the most advanced AI techniques for improving LLMs by integrating external knowledge sources~\cite{zhao2402retrieval}, ensuring reliability and providing up-to-date information. It offers significant convenience for a variety of tasks, including document assessment~\cite{hui2024uda}, financial market analysis~\cite{darji2024enhancing}, cybersecurity threat detection~\cite{munir2024leveraging}, and also science~\cite{su2024hybrid,shi2023retrieval}.

Although RAG is capable of retrieving and generating contextually relevant responses, the effectiveness of the outputs is largely influenced by the retriever's capacity to locate relevant and precise external resources. The unavoidable presence of noise, which often appears as irrelevant or misleading information, has the potential to introduce points of failure within RAG systems~\cite{barnett2024seven}.

% As an example, a recent investigation revealed that opting for top-k retrieval instead of just one enhances attribution but negatively affects the fluency in query-based RAG~\cite{aksitov2023characterizing}.

% 4) One of the main drawbacks of RAG, especially in its query-based form, is its significant increase in context length, which poses challenges for generators that have restricted context capabilities~\cite{liu2024lost}. Furthermore, the extended context tends to decelerate the overall generation process. Progress in the fields of prompt compression~\cite{jiang2023llmlingua} and long-context capabilities~\cite{han2023lm} has helped lessen these issues, although there is a minor sacrifice in terms of accuracy or expense.

Although RAG offers notable benefits in retrieving knowledge, its true potential is unlocked when it is paired with fine-tuning methods. This integration enables models to adjust and enhance their outputs according to particular task needs and specialised domain knowledge.

{\bf Fine tuning}
\newline
In the context of LLMs, fine-tuning adjusts the model parameters to improve performance on tasks such as classification, generation, or domain-specific reasoning, often yielding significant gains in accuracy and relevance~\cite{patil2024review}. The methods for fine-tuning LLMs vary from comprehensive approaches~\cite{hayou2024lora,li2025salora,he2025gora}, where all model parameters are adjusted, to more efficient strategies that update only a limited portion of parameters, thereby reducing computational overhead. 

However, when standard fine-tuning is applied to RAG, the interaction between the two methods is not as effective as anticipated, leading to additional challenges. The advantage of RAG lies in its ability to dynamically retrieve external knowledge, minimising the requirement for the model to store all relevant information within its parameters~\cite{soudani2024fine,lakatos2024investigating}. This tension undermines the flexibility of RAG, as the fine-tuned generator can prioritise its internalised knowledge over the retrieved context, leading to inconsistent or biased outputs.
% In safety-critical applications, fine-tuning might involve labeling software defect reports or safety compliance documents to tailor a model for precise assessment tasks~\cite{jo2025efficiency}.

%NVIDIA.~\cite{liu267035133chatqa} introduced ChatQA, a novel suite of models designed to surpass GPT-4 in both RAG and conversational QA. To improve generation, they proposed a two-stage instruction tuning method that significantly enhances RAG performance. In stage 1, supervised fine-tuning (SFT) is applied using a diverse mix of 128K samples from high-quality instruction-following and dialog datasets. Stage 2, called context-enhanced instruction tuning, integrates contextualized QA datasets—both human-annotated (HumanAnnotatedConvQA) and synthetic (SyntheticConvQA)—to boost the model’s ability to leverage retrieved or provided context. They also take retrieval models originally designed for single questions and fine-tune them to handle multi-turn conversations. This is one example of how fine tuning can be used in combination with RAG. However, we need a more customized solution specifically designed for assessing safety-critical software documentation, with significantly lower overhead and optimized for single-turn compliance queries on documentation. 

Although fine-tuning is effective in boosting performance, it carries notable trade-offs, requiring significant computational resources, high-quality labelled datasets, and careful expertise to prevent problems such as overfitting or catastrophic forgetting, where the model sacrifices its general knowledge as it specializes~\cite{zhang2024scaling,balaguer2024rag}. To overcome these challenges, researchers have explored hybrid approaches that integrate retrieval-based methods with fine-tuning~\cite{liu267035133chatqa}. In this direction, Balaguer et al.~\cite{balaguer2024rag} proposed a pipeline to combine both RAG and Fine-Tuning and analysed the trade-offs of each approach in several popular LLMs. Their approach involves fine-tuning an LLM based on RAG responses from Llama2-13B. Their findings indicate that fine-tuning improves model accuracy by more than 6\%, with RAG contributing an extra 5\% to the performance. This supports that fine-tuning LLMs in a RAG context likely improves performance. 

% For more effective retrieval, the authors presented a dense retriever optimized for conversational QA, yielding results on par with state-of-the-art query rewriting models while substantially reducing deployment costs. Additionally, they introduced ChatRAG Bench, a comprehensive evaluation suite comprising ten datasets that assess RAG, table-related QA, arithmetic calculations, and scenarios with unanswerable questions. 

% Balaguer et al.~\cite{balaguer2024rag} proposed a pipeline for combining both RAG and Fine-Tuning, and analyze the trade-offs of each approach across several popular LLMs, including Llama2-13B, GPT-3.5, and GPT-4. Their approach involves steps like retrieving data from PDFs, creating question-answer pairs, applying these for fine-tuning, and employing GPT-4 to assess outcomes. Their findings indicate that fine-tuning enhances model accuracy by more than six percentage points, with RAG contributing an extra five percentage points to performance improvements.

% When fine-tuning is applied to RAG, the interaction between the two methods is not as effective as anticipated, leading to additional challenges. The advantage of RAG lies in its capacity to dynamically retrieve external knowledge, thus minimizing the requirement for the model to store all relevant information within its parameters~\cite{soudani2024fine,lakatos2024investigating}. This tension undermines RAG’s flexibility, as the fine-tuned generator may prioritize its internalized knowledge over retrieved context, leading to inconsistent or biased outputs.

{\bf RAFT}

% The pretraining of LLMs on vast amounts of textual data has become a standard approach in NLP. However, when applying these pretrained models to downstream tasks, it is often necessary to integrate additional, domain-specific information. This integration is commonly achieved through methods like RAG prompting or fine-tuning, but determining the most effective method remains an unresolved issue. To address this issue, 
%The pretraining of LLMs on vast amounts of textual data has become a standard approach in NLP. However, when applying these pretrained models to downstream tasks, it is often necessary to integrate additional, domain-specific information. This integration is commonly achieved through methods like RAG prompting or fine-tuning, but determining the most effective way to combine them remains an open challenge.
Studies have shown that simply applying RAG and fine-tuning together does not necessarily lead to improved accuracy; in some cases, their interaction can even reduce performance. To address this issue, Zhang et al.~\cite{zhang2024raft} introduced Retrieval-Augmented Fine-Tuning (RAFT), a novel training technique designed to improve the model’s ability to answer questions in ``open-book'' in-domain settings. The key innovation of RAFT is its focus on training the model to ignore irrelevant information, or distractor documents, retrieved during the retrieval process. When training using RAFT, the model is guided to use only the relevant passages and cite them verbatim to help answer the question, improving reasoning capabilities, clarity and precision of responses. This approach, along with RAFT's chain-of-thought-style responses, significantly enhances performance and serves as an effective post-training method to improve pretrained LLMs when used with RAG. 

In our use case of safety-critical software assessment, RAFT presents a promising alternative to other conventional RAG and fine-tuning approaches. Since assessors must ensure that responses are not only contextually relevant but also traceable to authoritative documentation, RAFT's structured retrieval and fine-tuning process could enhance both the reliability and explainability of generated outputs. Building upon this, we have developed a technique called Document assessment Retrieval-Augmented Fine-Tuning (DRAFT), specifically tailored for the safety-critical software documentation domain, but generalisable to any document assessment application where there is a need to interrogate the document against domain-specific reference standards. DRAFT builds on the principles of RAFT but adapts the approach to address the unique requirements of safety-critical assessment contexts. Our methodology is designed to integrate with an existing specialised compliance assessment pipeline, with all design decisions informed by the specific demands of regulatory document assessment in high-assurance domains.

% Warnakulasuriya et al.~\cite{warnakulasuriya2024knowledge} proposed an innovative solution by integrating RAFT with LLMs to enhance first-aid training and its real-time application. By utilizing RAFT, AI models can access and incorporate vast medical knowledge bases, improving the accuracy and relevance of first-aid instructions in critical situations. 

% Building upon this, we have developed a technique called Document Retrieval-Augmented Fine-Tuning (DRAFT) specifically tailored for the safety-critical software documentation domain. DRAFT builds on the principles of RAFT but adapts the approach to focus on safety-critical tasks, ensuring that relevant documents are effectively retrieved and used to improve decision-making in high-stakes environments.

%%https://hyperight.com/7-practical-applications-of-rag-models-and-their-impact-on-society/%%

\section{Methodology}
\label{sec:methodology}
Decisions and motivations described in this section are based on state-of-the-art research. In this work, we addressed the compliance of operational technology cybersecurity (OTCS) documents by leveraging OTCS standards and documentation \cite{cyrail_paper}. We now extend this approach to the safety-critical software domain, where we face analogous compliance challenges. Our solution follows a similar methodology, replacing the OTCS standards with international standards for safety-critical software such as EN50716 and evaluating the documentation of safety-critical software using real case study data.

\subsection{RAG compliance pipeline}
\label{sec:methodology:rag_compliance_pipeline}

\begin{figure}[t]
  \centering
  \resizebox{\columnwidth}{!}{%
  \begin{tikzpicture}[node distance=1.2cm, every node/.style={fill=green!40, font=\sffamily}, align=center]
    \node (input) [draw, rounded corners] {Input\\Component};
    \node (prompt_template) [draw, rounded corners, right=of input] {Prompt\\Template};
    \node (dk_retriever) [draw, rounded corners, below=of input]{Context\\ Retriever};
    \node (ud_retriever) [draw, rounded corners, above=of input] {Document\\ Retriever};
    \node (llm) [draw, rounded corners, right=of prompt_template] {LLM};
    \node (output) [draw, rounded corners, right=of llm] {Output};
    \draw[->] (input) -- (dk_retriever);
    \draw[->] (input) -- (ud_retriever);
    \draw[->] (dk_retriever) -- (prompt_template);
    \draw[->] (ud_retriever) -- (prompt_template);
    \draw[->] (input) -- (prompt_template);
    \draw[->] (prompt_template) -- (llm);
    \draw[->] (llm) -- (output);
  \end{tikzpicture}
  }
  \caption{Flowchart of the compliance assessment pipeline}
  \label{fig_2}
\end{figure}

As shown in Figure~\ref{fig_2}, our compliance assessment pipeline implements a dual-retrieval architecture designed to improve compliance assessment queries on documentation. The system employs two concurrent retrievers:

(1) A document ($\mathcal{D}$) retriever that returns ($R_D(q)$) relevant information from user documentation based on the query $q$:
\begin{equation}
    R_D(q) = \{d_1, d_2, \ldots, d_n\} \quad \text{where } d_i \in \mathcal{D}
\end{equation}

(2) A context ($\mathcal{C}$) retriever that returns ($R_C(q)$) applicable standards and regulations based on the same query $q$:
\begin{equation}
    R_C(q) = \{c_1, c_2, \ldots, c_m\} \quad \text{where } c_i \in \mathcal{C}
\end{equation}

This parallel retrieval approach enables the LLM to process compliance questions more effectively by simultaneously providing domain-specific documentation and relevant regulatory context. By integrating these complementary knowledge sources, the architecture enhances the model's reasoning capabilities, resulting in more accurate and well-justified answers to compliance queries \cite{cyrail_paper}.

To optimise retrieval quality for both documents and standards, we implement a hybrid approach combining dense vector similarity and lexical matching. For dense retrieval, we employ top-$k$ similarity. For lexical matching, we implement BM25. We then linearly combine semantic and lexical relevance scores:

\begin{equation}
\begin{aligned}
    \text{score}_{\text{hybrid}}(q, x) &=  \alpha \cdot \text{sim}_{\text{dense}}(q, x) \\
    &\quad + (1 - \alpha) \cdot \text{norm}(\text{BM25}(q, x)) \\
    &\quad \text{where } \alpha = 0.75
\end{aligned}
\end{equation}

 % \begin{equation}
 % \begin{aligned}
 %     \text{score}_{\text{hybrid}}(q, x) &=  \alpha \cdot \text{sim}_{\text{dense}}(q, x) \\
 %     + (1 - \alpha) \cdot \text{norm}(\text{BM25}(q, x))
 %     \quad \\ \text{where } \alpha = 0.75
 % \end{aligned}
 % \end{equation}

The initial set retrieved for each retriever comprises the top-$10$ items ranked by this hybrid score. We then employ a reranking step using Cohere's reranker model\footnote{Information on Cohere Rerank 3.5 available here:\url{https://cohere.com/blog/rerank-3pt5}} to further refine these results, selecting only the $4$ most relevant chunks for the final retrieval set.

This two-stage retrieval process enables us to balance breadth and precision: first, capturing a wider set of potentially relevant chunks through embedding similarity and BM25, then refining this selection using Cohere's more computationally intensive but higher-quality reranking model.

The retrieved documents $R_D(q)$ and standards $R_C(q)$ are then combined within a structured prompt template and presented to the LLM, which generates the final answer to the compliance query:

\begin{equation}
\label{eq_2}
\begin{aligned}
    A = \text{LLM}(f_{\text{template}}(q, R_D(q), R_C(q)))
\end{aligned}
\end{equation}

where $A$ denotes the answer. To construct our retrieval corpus, we processed two distinct document collections. For the set of documents $\mathcal{D}$, we used internal case studies comprising real-world safety-critical software documentation. For the context set $\mathcal{C}$, we used the EN50716 safety-critical software standard\footnote{ENEN50716 standards accessible at: \url{https://knowledge.bsigroup.com}}%/products/railway-applications-communication-signalling-and-processing-systems-software-for-railway-control-and-protection-systems-1}}. %We applied a uniform processing pipeline to both collections:

%\begin{enumerate}
%    \item Document parsing: We converted PDFs to structured markdown using LlamaParse \cite{Liu2024}
%    \item Chunking: We segmented the documents using LlamaIndex's \cite{Liu_LlamaIndex_2022} sentence splitter with a chunk size of 512 tokens and an overlap of 20 tokens, resulting in page-sized information units
%    \item Embedding: We encoded all chunks using Cohere's embed-english-v3.0 model \cite{Cohere_2023}
%    \item Storage: We indexed the resulting embeddings in a vector database for efficient retrieval
%\end{enumerate}

%The resulting four chunks from each collection provide focused context in alignment with safety critical software assessment processes.

Despite our retriever enhancements, the pipeline comes with inherent limitations that we have aimed to resolve in this research. One of the conclusions of the previous work was that the retrieval system was imperfect, as are most retrievers in RAG applications \cite{gao2023retrieval,cyrail_paper}. Additionally, a significant limitation was identified that the LLM would often get confused between the two different categories of nodes, sometimes justifying the context as documentation due to their similarity. Furthermore, incorrect highly scored retrieval nodes would be falsely justified, resulting in lower correctness and poor reasoning. In response to these limitations, we aim to address these issues, improving our methodology by fine-tuning an LLM. Prior work has highlighted that, by teaching the LLM to ignore irrelevant chunks where retrieval has failed and showing the LLM how context and document chunks should be used in the answer, we can alleviate the limitations we have previously identified \cite{cyrail_paper}. %In addition to solving these problems, we also had some general aims for fine tuning such as the dataset creation being at least semi-automatic, ensuring that somewhere in the dataset there is human-in-the-loop and ensuring that the fine-tuning integrates with our multi-stage RAG pipeline and use case.

\subsection{Fine-tuning data}
\label{sec:methodology:ft-data}
%RAFT is a specialised technique to automatically create a fine tuning dataset for RAG pipelines to optimise Q\&A. 
%Given our requirements, we decided to create a process based on the retrieval augmented fine tuning (RAFT) paper \cite{zhang2024raft}.
%This approach boasts several benefits such as teaching the LLM not to use or justify irrelevant chunks in RAG, making it robust to the imperfect retrieval process. Using this well-tested approach, provides a good starting point for our fine tuning efforts as it directly tackles problems observed in our compliance pipeline.

Although effective for domain-specific question answering in standard RAG pipelines, RAFT cannot be directly applied to our use case and custom compliance pipeline. We face several integration issues that we must overcome:
\begin{enumerate}
    \item The dataset generation only includes a single category of node in its output dataset, whereas in our pipeline we have two categories of nodes, context and document.
    \item The automatic question generation in RAFT works to generally improve a QA use case. However, this technique cannot be automatically used to generate compliance questions that tailor to our use case.
    \item Compliance assessment queries can have positive and negative answers, that is, \texttt{complies because...}, \texttt{does not comply because...}. In RAFT there is always a single correct answer based on the context provided.
    \item In safety critical software assessments, assessors will likely look at multiple sections of documentation to form their answer; however, in the RAFT methodology the answers are generated based on a single chunk of context.% However, we have multiple chunks that we can use in our answer generation. 
    \item Finally, in our use case at least one document chunk is required in the prompt to answer the question whereas in RAFT this is not the case; ideally the context could be memorised from fine-tuning.% This means that we always have to include at least one golden document chunk in every fine tuning entry and the P parameter must be set to 0, otherwise the LLM might try to justify non-existent document chunks as it has learned to memorise them. 
\end{enumerate}

In order to help guide our future dataset generation process, we have defined the following 4 ``fine-tuning laws'' that will help with pipeline integration and ensure fine-tuning is a success:

\begin{enumerate}
    \item The fine-tuning dataset must contain examples of the task that we wish to improve during inference.
    \item The inputs and outputs of the dataset should reflect those of inference.
    \item The fine-tuning dataset should be varied and closely aligned to our use case.
    \item The dataset creation process should be semi-automatic and include our own data.
\end{enumerate}

Based on the above laws, it is clear that a fine-tuning dataset will look very similar to that of inference (see Equation \ref{eq_2}), hence we will need to define all these components in the context of our use case. To generate our list of compliance queries, $\mathcal{Q} = \{q_i \mid i \in \{1,2,\ldots,n\}\}$, we first collected a list of all the `shall' compliance statements from the EN50128 standards. Where large shall statements occurred, we split them into smaller shall statements. From these statements, we converted them into incomplete questions that could be prepended by ``does the user documentation contain''. For example, you shall do this $\to$ evidence that they do this $\to$ Does the user documentation contain evidence that they do this? We also added some additional questions based on internal guidance documents for safety-critical software assessment. In general, we collected 577 safety-critical software compliance questions that can be used in our fine-tuning dataset.

Our train, test and validation splits are 0.8, 0.1 and 0.1, respectively. Specifically, we divide $\mathcal{Q}$ and $\mathcal{D}$ as these are the factors that directly influence the answer. $\mathcal{Q}_{train}$ contained 465 questions $\mathcal{Q}_{test}$ and $\mathcal{Q}_{val}$ contained 56 questions each. For $\mathcal{D}$ we had access to 13 separate safety critical software projects totalling 9,220 pages. We decided to use 1 project for each $\mathcal{D}_{test}$ and $\mathcal{D}_{val}$, totalling 1,055 pages and 907 pages, respectively. The rest of the 11 projects are used for $\mathcal{D}_{train}$. The same $\mathcal{C}$ is used throughout training, testing and validation as preferably we would like this to be memorised. Wherever we describe using an LLM in our dataset generation process, we used OpenAI's GPT4o model \cite{hurst2024gpt}.

\subsection{Linking document chunks and compliance questions}

The original RAFT paper and code \cite{zhang2024raft} describe a technique to link the context chunk to the question to generate fine-tuning dataset entries (inputs only). Essentially, the technique involves asking an LLM to generate a question to which the context can be used as an answer. However, for our use case this does not work as we cannot reliably produce EN50716 compliant questions from just one document chunk. %Even including some context $R_C(d_i) = \{c_1, c_2, \ldots, c_m\} \quad \text{where } c_i \in \mathcal{C}$ in our question generation process does not seem rigorous and has the potential for hallucination.

%In Section \ref{sec:methodology:rag_compliance_pipeline} we described the creation of many different EN50128 compliant questions $\mathcal{Q}_{train}$. We considered the following approach:
%\begin{equation}
%\begin{aligned}
%D_i &= R_D(q_i) \\
%\mathcal{P} &= \{(q_i, D_i) \mid q_i \in \mathcal{Q}_{train}, D_i = R_D(q_i)\} \\
%&= \{(q_i, R_D(q_i)) \mid q_i \in \mathcal{Q}_{train}\} \\
%\text{where } &q_i \in \mathcal{Q}_{train}, \mathcal{Q}_{train} = \{q_1, q_2, \ldots, q_n\} 
%\end{aligned}
%\end{equation}

%Where:
%\begin{itemize}
%    \item $q_i$ represents a specific EN50128 compliant question from our predefined set
%    \item $R_D(q_i)$ is our retrieval function that finds document chunks relevant to question $q_i$
%   \item $D_i$ is the set of document chunks that are retrieved as relevant to question $q_i$
%    \item $\mathcal{P}$ represents the complete set of question-document pairs created by this reverse mapping
%    %\item $\mathcal{Q}_{train}$ is our complete set of predefined EN50128 compliant questions
%\end{itemize}

We considered an alternative approach of using our retriever $R_D$ to identify relevant document chunks for each compliance question in $\mathcal{Q}_{train}$, providing a link between $d_i$ and $q_i$. However, this approach has a significant limitation: it would only capture document chunks that the retriever deems relevant to our predefined questions, likely not including the full scope of our training data. We overcome this in the proposed final approach: %Ultimately, this approach would only use a subset of $\mathcal{D}_{train}$, therefore it is unlikely to capture the full distribution of our data. A more complicated approach in the spirit of Equation \ref{eq_3} looks like:
%\begin{equation}
%\begin{aligned}
%kQ_i &= R_Q(d_i) \\
%q_i^* &= LLM(d_i, kQ_i) \\
%\mathcal{P} &= \{(q_i^*, d_i) \mid d_i \in \mathcal{D}_{train}, q_i^* = LLM(d_i, kQ_i)\} \\
%%\text{where } &kQ_i \subset \mathcal{Q}_{train} \text{ is a set of relevant questions for } d_i \\
%%&q_i^* \in kQ_i \text{ is the most relevant question}\\
%\end{aligned}
%\end{equation}
\begin{equation}
\begin{aligned}
kQ_i &= R_Q(d_i) \\
S_i &= T(kQ_i) \\
q_i^* &= LLM(d_i, kQ_i, S_i) \\
\mathcal{P} &= \{(q_i^*, d_i) \mid d_i \in \mathcal{D}_{train}, q_i^* = LLM(d_i, kQ_i, S_i)\} \\
%\text{where } &kQ_i \subset \mathcal{Q}_{train} \text{ is a set of relevant questions for } d_i \\
%&S_i \text{ is the annex and section information extracted from references in } kQ_i \\
%&q_i^* \in kQ_i \text{ is the most relevant question}
\end{aligned}
\end{equation}
Where:
\begin{itemize}
    \item $R_Q(d_i)$ is our retrieval function that finds relevant questions for document chunk $d_i$.
    \item $kQ_i$ represents the top-k set of potentially relevant questions retrieved for document chunk $d_i$.
    %\item $LLM(d_i, kQ_i)$ is the process of using a large language model to select the optimal question from $kQ_i$ given document $d_i$
    \item $S_i$ is the annex and section information associated with references in $kQ_i$
    \item $T(kQ_i)$ is our pre-processing tool that extracts annex and section references and information from the candidate questions.
    \item $LLM(d_i, kQ_i, S_i)$ incorporates both the candidate questions and the information of the potentially associated section.
    \item $q_i^*$ is the single most relevant question selected by the LLM for document chunk $d_i$.
    \item $\mathcal{P}$ represents the set of question-document pairs.
\end{itemize}

The workflow can be summarised as follows.
\begin{enumerate}
    \item For each document chunk $d_i$, retrieve candidate questions $kQ_i$ using $R_Q(d_i)$.
    \item Process $kQ_i$ through our reference extraction tool $T$ to obtain context $S_i$.
    \item Provide the document chunk $d_i$, candidate questions $kQ_i$, and reference context $S_i$ to the LLM.
    \item The LLM selects the most appropriate question $q_i^*$ with full awareness of referenced content.
\end{enumerate}

In order to match each $d_i \in \mathcal{D}$ to a question $q_i \in \mathcal{Q}_{train}$ we first construct a retriever $R_Q(d_i)$ that stores each question from our train set as a chunk in a vector database. We use the same hybrid retrieval and reranking techniques described in Section \ref{sec:methodology:rag_compliance_pipeline}, except this time we return 25 questions from hybrid retrieval and rerank down to 5 questions.

An important consideration is that approximately 15\% of the questions contained information about section or annex information that is not explicitly known by the LLM. For example, a question entry might be ``Does the user documentation contain A Software Component Design Verification Report that has been written in accordance with the generic requirements established for a Verification Report (see 6.2.4.13)''. If we were to use any LLM-based matching of the document chunks to questions, then there would be insufficient context to accurately match a document to a question. We preprocess each question and extract all annexes and section references using a pattern matching tool. For each extracted reference, we create a dictionary that maps these references to the corresponding text blocks that contain the referenced information. When selecting the most appropriate question for a document chunk, we lookup this dictionary and include the relevant passages, $S_i$, as additional information for the LLM.
When the LLM makes the determination of which question is most relevant to a document chunk, it now has access to the complete context of the question, including any referenced sections or annexes that would otherwise be unknown.

\subsection{Grouping the dataset}

The question-document pairs $\mathcal{P} = \{(q^{*}_{i}, d_i)\}$ established in the previous section provide a foundational dataset for our approach. However, this simple pairing structure does not fully capture the complexity of real-world assessment scenarios. In Section \ref{sec:methodology:ft-data}, we establish that safety assessors in regulated industries often follow a ``multiple lines of evidence'' approach, where conclusions are drawn only after examining several related pieces of documentation. For example, to verify compliance with a specific safety requirement, an assessor might need to review design specifications, test results, and validation procedures collectively. Our fine-tuning process must reflect this reality to produce a model capable of handling such multi-document reasoning tasks effectively.

We implement a probabilistic document grouping strategy in which instead of maintaining strict one-to-one question-document pairs, we randomly group multiple document chunks that correspond to the same question. Formally, for each question $q^{*}_{i}$, we identify all matching document chunks $\{d_j | q^*_j = q^*_i\}$ and form random subsets of size $m$, where $1 \leq m \leq 4$. The parameter $m$ is randomly selected for each grouping to introduce variability in the training data.

In the RAFT paper \cite{zhang2024raft} the authors only experiment with a static number of golden chunks. Our approach offers several improvements. First, it creates a more diverse fine-tuning dataset that better represents real-world assessment scenarios. Second, it helps the model learn to synthesise information across multiple related documents. Third, and finally, it mitigates potential overfitting to single-document reasoning patterns. Given that $|\mathcal{D}_{train}| \gg |\mathcal{Q}_{train}|$, this grouping technique is effective and computationally feasible. 

An essential component of our inference pipeline is the context $\mathcal{C}$ retrieved through RAG mechanisms. For each grouped set of document chunks $\{d_{i1}, d_{i2}, \ldots, d_{im}\}$ associated with the question $q^*_i$, we retrieve relevant context chunks $c_i \in \mathcal{C}$ using our retrieval function $R_C(q^*_i)$. Specifically, we select the top $n$ context chunks, where $1 \leq n \leq 4$ and $n$ are randomly determined for each training instance. This randomisation in context size improves robustness to varying amounts of available contextual information,  creates additional variability in the training data, and completes a natural alignment with our inference pipeline (Equation \ref{eq_2}).

%The randomized selection of both $m$ and $n$ is further justified in Section \ref{sec_distractors}, where we discuss the impact of distractors on model performance.

%In our approach, we treat the set of document chunks $d_i$ and context chunks $c_i$ as "golden" or factual references to align with inference. %This modeling choice is motivated by our objective to create training conditions that closely mirror inference scenarios. During inference, the model must identify and extract relevant information from verified documentation. By treating grouped document chunks as factual ground truth during training, we align the learning objective with the inference task.

We now formally define our fine-tuning dataset $\mathcal{F}$ as:

\begin{equation}
\begin{aligned}
\mathcal{F} = \{(q^*_i, D\textasteriskcentered{}, C\textasteriskcentered{}) \mid q^*_i \in \mathcal{Q}_{train}, D\textasteriskcentered{} \subset \mathcal{D}_{train}, C\textasteriskcentered{} \subset \mathcal{C}\}
\end{aligned}
\end{equation}

Where:
\begin{itemize}
    \item $q^*_i$ is a question from our training question set.
    \item $D\textasteriskcentered{} = \{d_{i1}, d_{i2}, \ldots, d_{im}\}$ is a randomly sized subset of golden document chunks that all match to question $q^*_i$.
    \item $|D\textasteriskcentered{}| = m$ where $1 \leq m \leq 4$ is randomly selected.
    \item $C\textasteriskcentered{} = \{c_{i1}, c_{i2}, \ldots, c_{in}\}$ is a set of golden context chunks retrieved using $R_C(q^*_i)$.
    \item $|C\textasteriskcentered{}| = n$ where $1 \leq n \leq 4$ is randomly selected.
\end{itemize}

%This dataset construction procedure yields training instances that:
%\begin{enumerate}
%    \item Reflect real-world assessment scenarios requiring multiple document references
%    \item Incorporate varying amounts of relevant context
%    \item Maintain alignment between training and inference conditions
%    \item Provide sufficient variability to promote robust model learning
%\end{enumerate}

\subsection{Generating answers}

Having established our fine-tuning dataset $\mathcal{F}$ with questions, document groups, and context groups, we now turn to generating high-quality answers that leverage all available information.

\begin{figure*}
\begin{mdframed}
\footnotesize
\begin{verbatim}
You will be provided with some documentation and supporting context:
===================== **User Documentation**=====================
{user_docs_str}
=================================================================
--------------------- **Contextual Information** ----------------
{context_str}
-----------------------------------------------------------------
Based **solely** on the **User Documentation**
and by enhancing your analysis utilising the **Contextual Information**
please answer the following question.
**Question:** {query_str}
**Important Guidelines:**"
- **Do NOT** use any prior knowledge or external information."
- **Do NOT** perform an analysis of the **Contextual Information**
in your answer.
Your response **must** be in the following format:
- First Provide step-by-step reasoning on how to answer the **Question**,
potentially making use of the **Contextual Information**
to refine your steps,
do not directly mention **Contextual Information**.
- Explain which parts of the **User Documentation**
that are meaningful to answer the **Question** and explain why.
- Copy paste the relevant sentences from the **User Documentation**
in ##begin_quote## and ##end_quote##.
- Provide a summary of how you reached your answer.
\end{verbatim}
\end{mdframed}
\caption{Prompt template used to generate an answer in the fine tuning dataset}
\label{fig_3}
\end{figure*}

The prompt template in Figure \ref{fig_3} is designed to generate answers $a_i$ for each instance $(q^*_i, D\textasteriskcentered{}, C\textasteriskcentered{}) \in \mathcal{F}$, with each component serving a specific purpose:

\begin{itemize}
    \item \textbf{Information Hierarchy}: Establishes documentation as primary evidence while using contextual information as interpretive guidance.
    
    \item \textbf{Step-by-step Reasoning}: Implements a chain-of-thought approach to improve factuality and promote reasoning.
    
    \item \textbf{Evidence Identification}: By requiring explanation of relevant documentation parts, it teaches the model how documents are used in constructing the answer.
    
    \item \textbf{Direct Citation}: Mandating quotes creates explicit document-to-answer connections and enhances traceability.
    
    \item \textbf{Summarization}: Reinforces the reasoning path and conclusion.
\end{itemize}

This approach addresses the key challenges in our existing pipeline, as outlined in Section \ref{sec:methodology:rag_compliance_pipeline}. 

\subsection{Adding in distractors}

In real-world retrieval scenarios, RAG systems rarely return only relevant documents--they typically retrieve a mixture of relevant and irrelevant content. To create a fine-tuning dataset that better reflects this reality, we introduce the concept of ``distractors''—deliberately included irrelevant chunks that train the model to distinguish between useful and non-useful information. This approach builds on the RAFT methodology \cite{zhang2024raft}, which demonstrated significant accuracy improvements when including distractors in fine-tuning datasets for RAG systems.

For each instance in our fine-tuning dataset, we augment the fine-tuning dataset document chunks by adding a set of distractor document chunks. These distractors are selected from the remaining document pool $\mathcal{D}_{train} \setminus D\textasteriskcentered{}$, representing content that a retrieval system might incorrectly return as relevant but which does not directly contribute to answering the question $q^*_i$. Similarly, we introduce context distractors alongside the relevant context chunks $C\textasteriskcentered{}$.

To construct training instances that mimic real inference conditions, we define the expanded document and context sets used during fine-tuning:

\begin{equation}
\label{eq_3a}
D_{train}(q^*_i) = D\textasteriskcentered{} \cup D_{k}
\end{equation}

\begin{equation}
\label{eq_3b}
C_{train}(q^*_i) = C\textasteriskcentered{} \cup C_{k}
\end{equation}

Where:
\begin{itemize}
    \item $D\textasteriskcentered{}$ represents the set of $m$ golden document chunks directly relevant to answering query $q^*_i$, where $1 \leq m \leq 4$ as previously defined.
    \item $D_{k}$ represents the set of $(4-m)$ distractor document chunks sampled from $\mathcal{D}_{train} \setminus D\textasteriskcentered{}$.
    \item $C\textasteriskcentered{}$ represents the set of $n$ relevant context chunks that provide regulatory context for $q^*_i$, where $1 \leq n \leq 4$ as previously defined.
    \item $C_{k}$ represents the set of $(4-n)$ distractor context chunks sampled from $\mathcal{C} \setminus C\textasteriskcentered{}$.
\end{itemize}

This construction ensures that the total number of document chunks and context chunks presented to the model during training is fixed at 4 each. Critically, since the answer $q_i$ is generated exclusively from the golden chunks $D\textasteriskcentered{}$, the model explicitly learns to identify and ignore the distractor chunks. By training on this mixture, we develop the model's ability to distinguish between relevant and irrelevant information, which is one of our motivations for fine-tuning this pipeline.

Unlike the original RAFT experiments that used a fixed number of golden documents, our approach accommodates a variable number of golden documents ($1 \leq m \leq 4$). Although the RAFT methodology explores various configurations where the golden document is not included at all P\% of the time, our approach maintains at least one golden document ($m \geq 1$) in every training instance. For our use case, at least one authoritative source document is typically necessary to correctly answer a question. Furthermore, the optimal ``P value'' was irregular and only provided marginal performance gains in the RAFT paper; therefore, we decided it was not worth it to experiment changing the value.

%Our approach extends beyond standard RAFT implementations by incorporating variable numbers of golden documents ($1 \leq m \leq 4$) rather than a fixed count. This innovation addresses a limitation in the original RAFT study, which only tested static configurations (e.g., always exactly 1 golden document with n distractors). By varying the number of golden documents during training, our model develops the capacity to identify and integrate relevant information regardless of how much evidence is distributed across retrieved documents.

%This distractor-augmented fine-tuning framework creates a strong alignment between training and inference conditions. During inference, our retrieval system $R_D$ has the potential to return a mixture of relevant and irrelevant documents, with the model needing to identify and leverage only the relevant portions. By training on a similar distribution, we minimize the gap between training and inference environments and reinforce not justifying irrelevant document chunks and better identifying the relevant chunks.

Our final fine-tuning framework can be formalised as:

\begin{equation}
\begin{aligned}
\mathcal{F}= \{(q^*_i, D_{train}(q^*_i), C_{train}(q^*_i), a_i)\}
\end{aligned}
\end{equation}

Where $D_{train}(q^*_i)$ and $C_{train}(q^*_i)$ are as defined in Equations~\ref{eq_3a} and \ref{eq_3b}, consisting of a mixture of golden and distractor chunks.

According to our human-in-the-loop requirement, we decided to verify about 10\% of our training dataset using safety critical software assessors so that we could have assurance of the quality of our dataset. Approximately 30\% of the answers were modified in some way, and approximately 5\% of the answers required major modification. Due to the relatively low number of major modifications in our sample size, we deemed that it was not necessary to spend additional time modifying the rest of the training dataset. 

\subsection{Performing fine tuning}

\begin{figure}[t]
    \centering
    \scalebox{0.7}{
    \begin{tikzpicture}
        \begin{axis}[
            width=0.7\textwidth,
            height=0.35\textheight,
            xlabel={Record Number},
            ylabel={Loss Value},
            grid=both,
            title={4o-mini Model: Complete Training and Validation Loss Progression (All 685 Records)},
            xmin=0, xmax=700,
            ymin=0, ymax=1.2,
            legend pos=north east,
            ]
            
            % All training loss measurements, sampled every 5 points to prevent memory issues
            \addplot[
                blue,
                mark=none,
                thin,
                smooth,
            ] coordinates {
                (1, 1.0049) (5, 0.7948) (10, 0.9720) (15, 0.7921) (20, 0.8009)
                (25, 0.6372) (30, 0.5831) (35, 0.5207) (40, 0.5386) (45, 0.3889)
                (50, 0.4354) (55, 0.4926) (60, 0.3768) (65, 0.3300) (70, 0.3212)
                (75, 0.3997) (80, 0.3909) (85, 0.3524) (90, 0.3026) (95, 0.2472)
                (100, 0.2938) (105, 0.2926) (110, 0.3194) (115, 0.3601) (120, 0.2599)
                (125, 0.2238) (130, 0.2954) (135, 0.1974) (140, 0.2413) (145, 0.1947)
                (150, 0.2167) (155, 0.2862) (160, 0.1958) (165, 0.1760) (170, 0.2241)
                (175, 0.3105) (180, 0.2133) (185, 0.2137) (190, 0.2430) (195, 0.2030)
                (200, 0.1997) (205, 0.2394) (210, 0.2720) (215, 0.1874) (220, 0.2909)
                (225, 0.3090) (230, 0.2548) (235, 0.2693) (240, 0.3343) (245, 0.2057)
                (250, 0.1766) (255, 0.2649) (260, 0.2033) (265, 0.2597) (270, 0.2317)
                (275, 0.2035) (280, 0.2207) (285, 0.2154) (290, 0.2465) (295, 0.2847)
                (300, 0.2332) (305, 0.1908) (310, 0.2704) (315, 0.2772) (320, 0.2367)
                (325, 0.2053) (330, 0.1860) (335, 0.2187) (340, 0.1962) (345, 0.2159)
                (350, 0.1391) (355, 0.2831) (360, 0.2336) (365, 0.1834) (370, 0.2681)
                (375, 0.2147) (380, 0.2412) (385, 0.2094) (390, 0.2170) (395, 0.2025)
                (400, 0.2819) (405, 0.2246) (410, 0.2124) (415, 0.2782) (420, 0.2605)
                (425, 0.2556) (430, 0.2061) (435, 0.2991) (440, 0.2388) (445, 0.2603)
                (450, 0.2830) (455, 0.1907) (460, 0.2783) (465, 0.1899) (470, 0.1715)
                (475, 0.2726) (480, 0.2310) (485, 0.1760) (490, 0.1912) (495, 0.2045)
                (500, 0.2003) (505, 0.1866) (510, 0.1976) (515, 0.1888) (520, 0.1961)
                (525, 0.1737) (530, 0.2526) (535, 0.1893) (540, 0.3016) (545, 0.2223)
                (550, 0.1756) (555, 0.2194) (560, 0.2318) (565, 0.2273) (570, 0.2271)
                (575, 0.2181) (580, 0.3161) (585, 0.1975) (590, 0.2777) (595, 0.2174)
                (600, 0.2206) (605, 0.1955) (610, 0.2062) (615, 0.2044) (620, 0.2693)
                (625, 0.2115) (630, 0.1746) (635, 0.1768) (640, 0.2413) (645, 0.1787)
                (650, 0.2400) (655, 0.2050) (660, 0.2607) (665, 0.2497) (670, 0.1866)
                (675, 0.2244) (680, 0.1992) (685, 0.2722)
            };
            \addlegendentry{Training Loss}
            
            % Validation loss measurements as a line
            \addplot[
                red,
                mark=*,
                mark size=1.5pt,
                thick,
            ] coordinates {
                (10, 0.8937) (20, 0.6662) (30, 0.6922) (40, 0.5856) (50, 0.4568)
                (60, 0.3518) (70, 0.3461) (80, 0.3021) (90, 0.2793) (100, 0.3061)
                (110, 0.2647) (120, 0.2155) (130, 0.2853) (140, 0.2626) (150, 0.2191)
                (160, 0.2245) (170, 0.1840) (180, 0.2483) (190, 0.1795) (200, 0.2335)
                (210, 0.2568) (220, 0.2313) (230, 0.2614) (240, 0.2022) (250, 0.2242)
                (260, 0.1653) (270, 0.2964) (280, 0.2850) (290, 0.2196) (300, 0.2526)
                (310, 0.2150) (320, 0.2037) (330, 0.2429) (340, 0.2316) (350, 0.1973)
                (360, 0.2408) (370, 0.1987) (380, 0.1767) (390, 0.2302) (400, 0.2289)
                (410, 0.1764) (420, 0.2430) (430, 0.2554) (440, 0.2102) (450, 0.2022)
                (460, 0.1851) (470, 0.2269) (480, 0.2442) (490, 0.2166) (500, 0.2369)
                (510, 0.1865) (520, 0.2032) (530, 0.2708) (540, 0.2528) (550, 0.1946)
                (560, 0.2010) (570, 0.2312) (580, 0.1874) (590, 0.2253) (600, 0.2249)
                (610, 0.2377) (620, 0.2378) (630, 0.1773) (640, 0.2034) (650, 0.2418)
                (660, 0.1656) (670, 0.1925) (680, 0.2413) (685, 0.2906)
            };
            \addlegendentry{Validation Loss}
            
            % Vertical lines at key points
            \draw[dashed, gray] (axis cs:100,0) -- (axis cs:100,1.2);
            \draw[dashed, gray] (axis cs:300,0) -- (axis cs:300,1.2);
            \draw[dashed, gray] (axis cs:500,0) -- (axis cs:500,1.2);
            
            % Annotations
            \node[anchor=south, font=\small] at (axis cs:100,1.1) {Record 100};
            \node[anchor=south, font=\small] at (axis cs:300,1.1) {Record 300};
            \node[anchor=south, font=\small] at (axis cs:500,1.1) {Record 500};
            
        \end{axis}
    \end{tikzpicture}
    }
    \caption{4o-mini model: Full visualization of training and validation loss across all 685 records. The blue line shows the training loss (sampled every 5 points for clarity), while the red line with markers shows validation loss measurements taken at every 10th record. Note the significant decline in both losses during the first 100 records and the stabilization after approximately record 300.}
\label{fig-4o-mini-loss}
\end{figure}
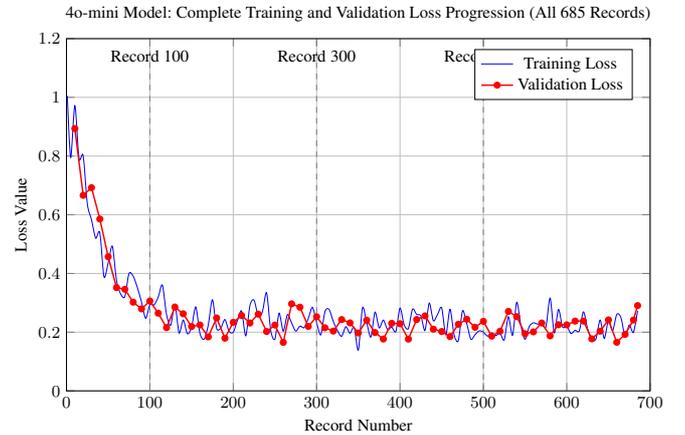

We used Low-Rank Adaptation (LoRA) \cite{hu2022lora} for fine-tuning both models. LoRA offers significant advantages over full fine-tuning, particularly for large language models. By decomposing weight updates into low-rank matrices, LoRA dramatically reduces the number of trainable parameters while maintaining performance comparable to full fine-tuning. This approach is especially beneficial in our safety assessment context, where deployment efficiency and resource constraints are important considerations. Additionally, LoRA has been shown to reduce the risk of catastrophic forgetting \cite{aghajanyan2020intrinsic}, helping the model retain its general capabilities while adapting to our specialised task.

For our fine-tuning experiments, we decided to fine-tune 4o-mini \cite{hurst2024gpt}. Our reasoning for selecting 4o-mini and not an even smaller model was that it is likely that the non-fine-tuned model would not produce comprehensive answers due to the complexity of the use case. This means that comparing the two models would be ineffective.

\begin{table}[t]
\centering
\begin{tabular}{l|c}
\hline
\textbf{Hyperparameter} & \textbf{Value} \\
\hline
Training dataset size & 3,422 entries \\
Validation dataset size & 342 entries \\
Trained tokens & 9,777,941 \\
Epochs & 1 \\
Batch size & 4 \\
Learning rate multiplier & 0.2 \\
\hline
\end{tabular}
\caption{Fine-tuning hyperparameters for the 4o-mini models}
\label{tab:hyperparams}
\end{table}
For fine-tuning we used the following hyperparameters in Table \ref{tab:hyperparams}.
Figure~\ref{fig-4o-mini-loss} presents the training and validation loss trajectories. 

%Both models demonstrate similar learning patterns, with rapid improvement during the first 100-150 records, followed by more gradual improvements and eventual stabilization. This consistent pattern across both model sizes suggests that our fine-tuning approach effectively teaches the models to perform our specialized task regardless of model scale.

We observe that the training loss decreases from approximately 1.0 to stabilise around 0.2, while the validation loss follows a similar trajectory from about 0.9 to 0.2. The close alignment between training and validation loss curves indicates that the model generalises well to unseen data rather than simply memorising the training examples by overfitting.

%The 4o model exhibits slightly lower initial loss values (starting at approximately 0.76 versus 1.0 for 4o-mini) and converges slightly faster to similar final loss values around 0.2. This aligns with expectations that larger models often have an initial advantage due to their ability to understand more complex tasks.

%Importantly, neither model shows signs of overfitting, as evidenced by the validation loss tracking closely with the training loss throughout the fine-tuning process. This suggests our dataset size, architectural choices, and hyperparameters strike an appropriate balance for the task complexity.

The stabilisation of the loss after approximately 300 records indicates that the model has reached a point of diminishing returns in learning from additional examples. This is particularly encouraging given that our hyperparameters included only a single epoch, suggesting efficient learning without the need for multiple passes through the dataset. Given the final loss values and the convergence patterns observed, we consider the fine-tuning to be successful.

\section{Experiment and Results}
\label{sec:results}

\subsection{Experiment details}
To evaluate the quality of the response in the safety critical software domain, we blindly assessed our test set using a 0-10 correctness metric based on the conformity to ideal responses. The evaluations were conducted by a safety critical software assessor on a set of 56 questions, the results of which are presented in Table \ref{tab}.

\subsection{Results}
\begin{table}[t]
\centering
\begin{tabular}{l|c}
\hline
\textbf{Model} & \textbf{Avg Rating} \\
\hline
4o-mini & 6.50 \\
\hline
ft-4o-mini & 6.98 \\
\hline
\end{tabular}
\caption{Comparison of 4o-mini base model vs. fine-tuned version}
\label{tab}
\end{table}
The fine-tuned model demonstrated a 7\% increase in performance compared to the base model, suggesting moderate improvements in response quality.

\section{Discussion}
\label{sec:discussion}
Our analysis revealed several key differences between the base and fine-tuned models that explain the modest performance improvement observed. These differences fall into three main categories: evidence handling, response structure, and domain understanding.

Regarding evidence handling, the fine-tuned model demonstrated improved precision in document use. It consistently avoided referencing irrelevant documents, unlike the base model, which often explained document chunks regardless of relevance. This targeted approach significantly improved response clarity and conciseness. However, the fine-tuned model occasionally over-relied on direct evidence, sometimes hesitating to make reasonable assumptions when documentation was incomplete.

In terms of response structure, the fine-tuned model produced more well-organised answers with clearer examples supporting its reasoning. Its justifications were generally more convincing and detailed, particularly beneficial for complex questions requiring depth. This verbosity, while advantageous for nuanced queries, occasionally introduced unnecessary complexity for simpler questions.

The fine-tuned model's enhanced domain understanding was evident in its ability to recognise relevant evidence that the base model was overlooked due to insufficient safety-critical software knowledge. This specialised expertise allowed the fine-tuned model to demonstrate greater confidence in its responses.

The relatively modest 7\% improvement suggests that while fine-tuning provided clear benefits in specific areas, the base model already performed reasonably well in this domain. The improvements were qualitative rather than transformative, with the fine-tuned model excelling in precision and domain-specific understanding while sometimes sacrificing flexibility. These findings indicate that targeted fine-tuning offers measurable but incremental improvements for specialised applications in safety critical software contexts.

\section{Conclusion}
\label{sec:conclusion}

In this paper, we present Document Retrieval-Augmented Fine-Tuning (DRAFT), a novel approach to enhance LLM performance in safety-critical software assessment tasks. Our results demonstrate a modest but meaningful 7\% improvement in correctness, though this metric fails to capture the more subjective enhancements observed across the varying types of questions. The improvement was particularly pronounced for complex queries that required domain expertise, while simpler questions showed less dramatic gains.

Several limitations should be acknowledged. Unlike classification tasks, evaluating compliance assessment responses is inherently subjective and requires expert human evaluation. This introduces potential variability despite our efforts to standardise assessment criteria. Furthermore, the high baseline performance of modern LLMs like GPT-4o-mini may create a ceiling effect that limits the observable impact of fine-tuning.

Future work could explore several promising directions. First, systematic hyperparameter optimisation might yield further performance gains, particularly in areas such as distractor ratios and learning rate. Second, applying our methodology to different contexts within the engineering domain would test its generalisability. Third, experimenting with alternative prompt templates for answer generation could further enhance the model's ability to recognise and utilise domain-specific information.

Our findings raise broader questions about the value proposition of fine-tuning for RAG systems. The modest performance gains observed may suggest that, for organisations with access to state-of-the-art LLMs, the additional investment in fine-tuning might not always be justified. Larger models may inherently possess sufficient reasoning capabilities to handle complex compliance tasks effectively without specialised training. Nevertheless, for resource-constrained environments or specialised applications, our DRAFT approach offers a viable path to enhancing domain-specific capabilities while maintaining evidence-based reasoning that is critical in safety critical software assessment contexts.

% \begin{ack}
% By using the \texttt{ack} environment to insert your (optional) 
% acknowledgements, you can ensure that the text is suppressed whenever 
% you use the \texttt{doubleblind} option. In the final version, 
% acknowledgements may be included on the extra page intended for references.
% \end{ack}
%\bibliographystyle{IEEEtran}
\bibliography{bib}

\end{document}